# Exchange interactions and Curie temperature of Ce-substituted SmCo5

S.-Y. Jekal[1, 2, *]

[1]*Laboratory of Metal Physics and Technology, Department of Materials, ETH Zurich, 8093 Zurich, Switzerland*
[2]*Condensed Matter Theory Group, Paul Scherrer Institute, CH-5232 Villigen PSI, Switzerland*
(Dated: August 2, 2018)

A partial substitution of Sm by Ce can have drastic effects on the magnetic performance, because it will introduce strain in the structure and breaks the lattice symmetry in a way that enhances the contribution of the Co atoms to magnetocrystalline anisotropy. However, Ce substitutions, which are benefit to improve the magnetocrystalline anisotropy, are detrimental to enhance the Curie temperature ($T_C$). With the requirements of wide operating temperature range of magnetic devices, it is important to quantitatively explore the relationship between the $T_C$ and ferromagnetic exchange energy. In this paper we show, based on mean-field approximation, that Ce substitution-induced tensile strain in SmCo5 leads to enhanced effective ferromagnetic exchange energy and $T_C$ while Ce atom itself reduces $T_C$.

## I. INTRODUCTION

Owing to the large magnetocrystalline anisotropy, high Curie temperature ($T_C$) and saturation magnetization ($M_s$), Sm–Co compounds have drawn attention for the high-performance magnet[1–12]. The net performance of the magnet crucially depends on the inter-atomic interactions, which in turn depend on the local atomic arrangements. In order to improve the intrinsic magnetic properties of Sm–Co system, further approaches[5,7,9,12–19] have been attempted in the past few decades.

A promising scheme is the partial substitution of Co with other transition metals (TM), such as Cu, Ni, Fe, or Zr in Sm(Co, TM)5[15,19–22]. The partial TM substitution enhances magnetic anisotropy energy (MAE) due to slight modifications of the atomic structure and the breaking of the lattice symmetry. In most cases, however, due to the reduced exchange coupling parameters, both $T_C$ and the $M_s$ of the Sm(Co, TM)5 compounds decrease with increasing TM content.[15,19,20]

Importantly, any substitution of Co with another TM reduce the effective ferromagnetic exchange coupling and consequently reduce $T_C$. This is highly undesirable because the most important application field of Sm-Co-based magnets is that of high temperature. Hence, it is vital that we find a way to enhance the magnetic performance in Sm-Co magnets without substituting away the Co, but by carefully modifying the local atomic arrangements, by introducing strain. Strain is a particularly important issue for permanent magnets, because their synthesis typically involves sintering, and thermal processes inevitably introduce strain on fine-grained materials[23,24]. Strain effects are also associated directly to partial element substitution, such as e.g. of Sm with another RE metal.

Ce is a particularly promising candidate because it is the most abundant and second lightest among all RE metals, and CeCo5 is isomorphous with SmCo5[25,26]. It is well-known that the magnetic material parameters of CeCo5 are substantially weaker than those of SmCo5[27], i.e., reduced $T_C$ of 660 K instead of 1000 K, reduced

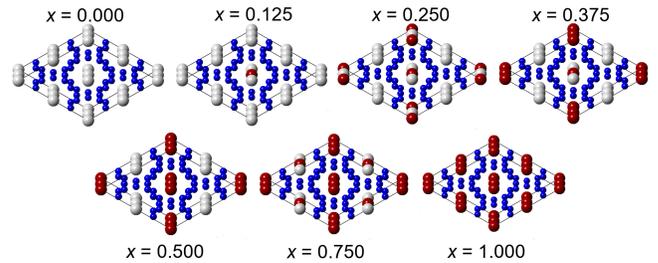

FIG. 1. Illustration of the crystal structure of $Sm_{1-x}Ce_xCo_5$ for $x$=0, 0.125, 0.25, 0.375, 0.5, 0.75, and 1. Gray, red, and blue spheres correspond to Sm, Ce, and Co atoms, respectively.

MAE of 5.3 MJ/m$^3$ instead of 17.2 MJ/m$^3$, and reduced $M_s$ of 0.95 T instead of 1.15 T. Nevertheless, in recent, we found a partial substitution of Sm by Ce can have drastic effects on the magnetic performance, because it will introduce strain in the structure and breaks the lattice symmetry in a way that enhances the contribution of the Co atoms to magnetocrystalline anisotropy (MCA)[28]. However, Ce substitutions, which are benefit to improve the MCA with slight modifications of the local atomic arrangements, are detrimental to enhance the Curie temperature. With the requirements of wide operating temperature range of magnetic devices, it is very important to quantitatively explore the relationship between the Curie temperature $T_C$ and ferromagnetic exchange energy.

In this paper we will report, based on mean-field approximation (MFA), that Ce substitution-induced tensile strain in SmCo5 leads to enhanced effective ferromagnetic exchange energy and $T_C$, however Ce atom itself reduces $T_C$.

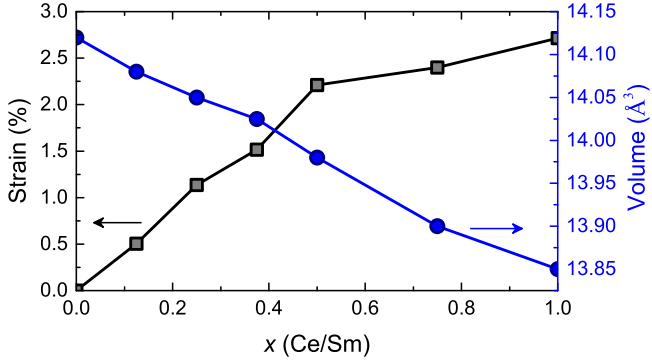

FIG. 2. Modification of the effective strain (square) compared to the SmCo$_5$ cell, and the atomic volume (circle) in Sm$_{1-x}$Ce$_x$Co$_5$, as a function of Ce substitution $x$.

## II. RESULTS AND DISCUSSION

### A. Crystal Structure

The intermetallic magnetic compounds SmCo$_5$ and CeCo$_5$ crystallize into the hexagonal CaCu$_5$ structure, where each Sm (Ce) occupies a $1a$ site and Co occupies $2c$ and $3g$ sites in the lattice that Co$_3$ and SmCo$_2$ sub-layers are alternatively along the $c$-axis[29]. To understand the Ce substitution effect on crystal structure and magnetic properties, we performed calculations of the Sm$_{1-x}$Ce$_x$Co$_5$ ($x = $ Ce/Sm) system, where we considered seven different compositions: $x$=0, 0.125, 0.25, 0.375, 0.5, 0.75, and 1. Configurations shown in Fig. 1 are the most stable structures obtained by total energy minimization in the density functional theory calculations[28].

The atomic volume decreases linearly from 14.12 Å$^3$ in SmCo$_5$ to 13.84 Å$^3$ in CeCo$_5$ with increasing $x$, which is due to the smaller atomic radius of Ce compared to that of Sm, whereas the strain increases correspondingly monotonically with increasing $x$ (see Fig. 2). It has been known that substitution-induced tensile strain on Sm$_{1-x}$Ce$_x$Co$_5$ has drastic effects on the magnetic state[28].

### B. Exchange Interactions

Mean-field Hamiltonian for the 3-sublattice system could be written as

$$H_{MFA} = -J_{c-c}m_c \sum_i^{N_c} S_{c,i} - 2J_{c-g}m_c \sum_i^{N_g+N_c} S_{g,i} - J_{g-g}m_g \sum_i^{N_g} S_{g,i} - 2J_{g-a}m_g \sum_i^{N_g+N_a} \sigma_{a,i} - 2J_{c-a}m_c \sum_i^{N_c+N_a} \sigma_{a,i},$$

(1)

where $J_{ij}$, $m_i$, and $S_i$ are exchange coupling between sites $i$ and $j$, magnetization, and classical spin vector at crystal site $i$, respectively. Labels of $c$ and $g$ correspond to $2c$ and $3g$ sites of Co atom, while a represents $1a$ site of Sm.

Sublattice magnetization is

$$m_i = \frac{2S_i+1}{2S_i} \coth(\frac{2S_i+1}{2S_i} \frac{1}{T} \sum_{ij} z_{ij} J_{i-j} m_j) - \frac{1}{2S_i} \coth(\frac{1}{2S_i} \frac{1}{T} \Sigma_{ij} z_{ij} J_{i-j} m_j)$$

(2)

where $z_{ij}$ is the number of neighboring sites in $j$th sublattice to $i$th sublattice.

While the inter-atomic ferromagnetic exchange interactions can be obtained as the difference between ferromagnetic and antiferromagnetic spin configuration, i.e., $J = (E_{\uparrow\uparrow} - E_{\uparrow\downarrow})/2$ by using density functional theory calculations[28], the effective exchange energies of Co and Sm are calculated as

$$J_{\text{eff}}^{\text{Co}} = (z_{gg} + z_{gc} + z_{cc} + z_{cg})J_{\text{Co-Co}}$$
$$J_{\text{eff}}^{\text{R}} = (z_{ac} + z_{ca} + z_{ag} + z_{ga})J_{\text{Co-R}}$$

(3)

, respectively.

Due to the smallness of the de Gennes factor of Sm,

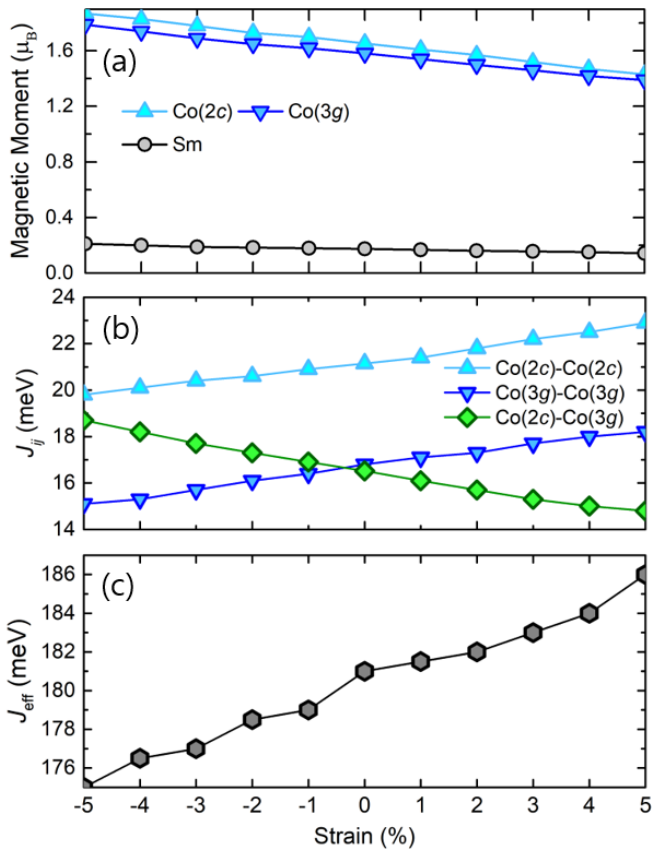

FIG. 3. Intrinsic magnetic properties as a function of strain in SmCo₅: (a) atomic magnetic moments of Co atoms at sites 2c (triangles) and 3g (inverse-triangles) and Sm atoms (circles); (b) exchange coupling between Co sites $i$ and $j$; and (c) effective total ferromagnetic exchange energy.

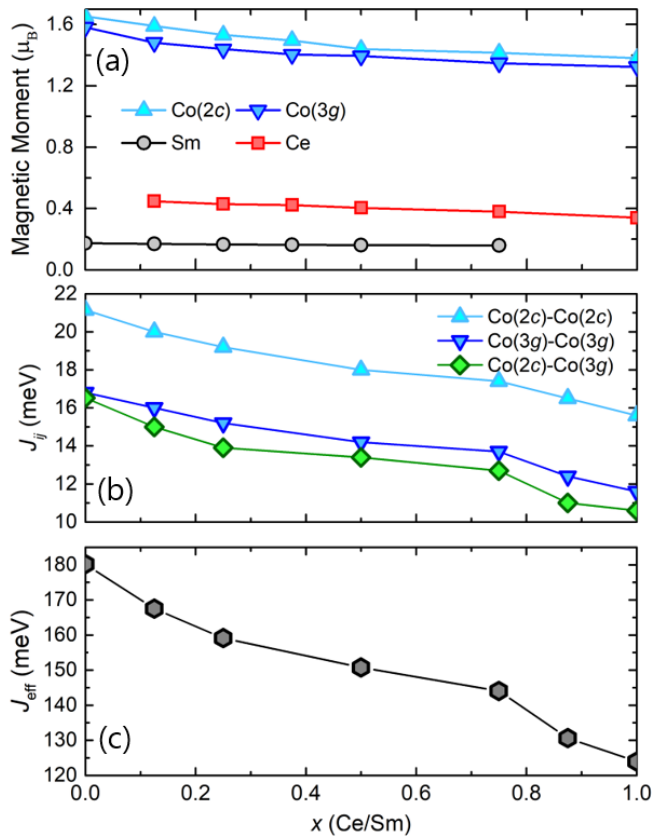

FIG. 4. Magnetic properties as a function of Ce-substitution of Sm: (a) atomic magnetic moments of Co atoms at sites 2c (triangles) and 3g (inverse-triangles) and Sm atoms (circles); (b) exchange coupling between Co sites $i$ and $j$; (c) effective total ferromagnetic exchange energy.

the Curie temperature is dominantly determined by the interatomic exchange interaction of the Co atoms[9,30]. Therefore the calculations on the exchange interaction are focused on the Co-Co couplings. To understand the role of Ce substitution-induced strain and Ce substitution itself, magnetic moment, inter-site exchange coupling ($J_{ij}$) and effective exchange energy ($J_{eff}$) are calculated as a function of the strain in SmCo₅ (Fig. 3) and Ce concentration in Sm$_{1-x}$Ce$_x$Co₅ (Fig. 4).

In Fig. 3(a), the magnetic moments at each crystallographic site decrease nearly linearly with increasing tensile strain. The reason for the decrease of the magnetic moments is the decreasing overlap of the $d$-orbitals as the $c$-axis grows with strain. Further, we calculate the inter-site exchange coupling and overall effective exchange coupling parameters. The intra-plane interaction strengths $J_{Co(2c)-Co(2c)}$ and $J_{Co(3g)-Co(3g)}$ increase while the inter-plane interaction strength $J_{Co(2c)-Co(3g)}$ decreases with increasing tensile strain (Fig. 3b). This illustrates the dependence of the ferromagnetic exchange on the inter-atomic distance: as the $c$–$c$ and $g$–$g$ distances decrease the corresponding exchange interaction increases, whereas as the c–g distance increases with increasing tensile strain the exchange interaction decreases. Then we obtain effective exchange interactions $J_{eff}$, which are the sum of the exchange coupling constants within a sphere of radius $R = 5a$. From these calculations we observe that the total ferromagnetic exchange interaction increases monotonically with increasing strain (see Fig. 3c), in contrast to the magnetic moments.

For Ce substituted Sm$_{1-x}$Ce$_x$Co₅, the calculated magnetic moments of the Co, Ce, and Sm atoms decrease monotonically with increasing Ce-substitution $x$, which correlates to the decreasing atomic volume, following Vergard's law[31] (see Fig. 4a). Also, the calculated $J_{Co-Co}$ and $J_{eff}$ decrease strongly with increasing $x$, as seen in Fig. 4(b) and (c).

## C. Calculations of Curie temperature

The corresponding Curie temperatures are also derived from the calculated exchange parameters based on the MFA.

The MFA is based on the notion of single-spin excita-tions, and the Hamiltonian is

$$\hat{\mathbf{h}}_i = -\vec{\mu}_i \cdot \vec{H}_i = -g\mu_B \vec{S}_i \cdot \vec{H}_i \qquad (4)$$

where $g$ and $\mu_B$ are the Lande factor and Bohr magneton, respectively. The molecular field could be thus defined as

$$\hat{\mathbf{H}}_i = -\frac{1}{g\mu_B} \sum_j J_{ij} < \vec{S}_j > \qquad (5)$$

with

$$< S_{iz} > = S_i B_{S_i}(x_i), x_i = \frac{g\mu_B H_i}{k_B T} \qquad (6)$$

where $B_{S_i}(x_i)$ is the Brillouin function, $k_B$ is the Boltzmann constant, and $T$ is the temperature in K. When $T$ is very high, such as $k_B T >> \mu_B H_i$, the $B_{S_i}(x_i)$ and $< S_{iz} >$ are rewritten as

$$B_s(x) = \frac{(S+1)x_i}{3}, < S_{iz} > = \frac{S_i(S_i)\mu_B H_i}{3k_B T}. \qquad (7)$$

By using the exchange integral $J_{ij}$ previously obtained, we get

$$< S_{iz} > = -\frac{S_i(S_i+1)}{3k_B T} \sum_j J_{ij} < S_{jz} >,$$
$$T < S_{jz} > + \frac{S_i(S_i+1)}{3k_B} \sum_j J_{ij} < S_{jz} > = 0 \qquad (8)$$

which has nonzero solution only if the determinant

$$\begin{bmatrix} a_{11} - T & \dots & \dots & a_{1n} \\ \dots & \dots & \dots & \dots \\ a_{n1} & a_{n2} & \dots & a_{nn} = T \end{bmatrix} = 0, a_{ij} = \frac{S_i(S_i+1)}{3k_B} \Sigma_j J_{ij}. \qquad (9)$$

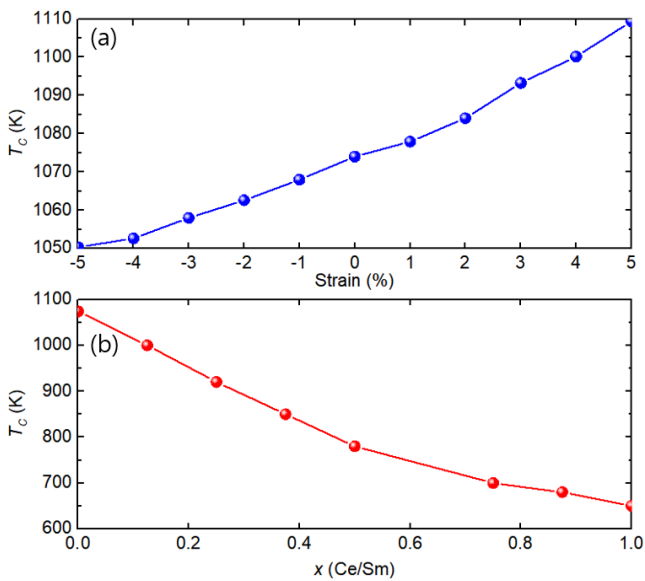

FIG. 5. Variation of the Curie temperature with respect to (a) strain and (b) Ce concentration $x$.

Among solutions in Eq. (9), the highest positive $T$ is the desired Curie temperature $T_C$.

As shown in Fig. 5(a), $T_C$ increases from 1085 K to 1110 K with increasing tensile strain along $c$-axis from 0% to 5%. For non-strained SmCo$_5$ (0%), the calculated $T_C$ (1085 K) is comparable with the experimental values ($T_C$ = 1020 K)[8], considering the MFA Curie temperature is generally overestimated by about 10-20%[32]. On the other hand, $T_C$ rapidly reduces with increasing concentration of Ce substitutions, and eventually CeCo$_5$ exhibit only $T_C$=660 K. Because for Sm$_{1-x}$Ce$_x$Co$_5$ the decrease of the effective ferromagnetic exchange is reinforced by the decrease of the magnetic moments. The appearance of the value of $T_C$ results from the strain dependence of the effective exchange parameters $J_{eff}$.

## III. CONCLUSIONS

Using mean-field theory approximation, we investigated the Ce substitution and strain effect on ferromagnetic exchange energy and corresponding Curie temperature of SmCo$_5$ compounds. It is found that tensile strain along $c$-direction improves $T_C$ which is key property for hard magnets. In comparison with an equilibrium state of SmCo$_5$, about 40 K higher $T_C$ is observed when 5% tensile strain is applied while Ce substitutions rapidly reduce the $T_C$. The enhancement and reduction of $T_C$ can be explained by responses of exchange energy parameters to the lattice strain and Ce substitution, respectively.

## IV. ACKNOWLEDGMENTS

The author gratefully acknowledge funding from the ETH Grant ETH-47 17-1.